\documentclass[12pt,a4paper]{article}
\begin{document}
\textwidth=135mm
 \textheight=200mm
\begin{center}
{\bfseries Two-pion interferometry for granular sources}
\vskip 5mm
Wei-Ning Zhang \vskip 5mm {\small {\it School of Physics and
Optoelectronic Technology, Dalian University of
Technology, Dalian, 116024, China}} \\
\end{center}
\vskip 5mm \centerline{\bf Abstract} A review on the two-pion
Hanbury-Brown-Twiss (HBT) interferometry in the granular source
model of quark-gluon plasma droplets is presented. The
characteristic quantities of the granular source extracted by
imaging analysis are presented and compared with the HBT radii
obtained by the usual Gaussian formula fit. The signals of granular
sources are presented. \vskip 10mm
\section{\label{sec:intro}Introduction}
The main purpose of relativistic heavy ion collisions is the study
of the quark-gluon plasma (QGP) formed in the early stage of the
collisions. As is well known, the system produced in the collisions
may thermalize and reach local equilibrium in a very short time
$\tau_0$. The following evolution can be described by hydrodynamics.
Hydrodynamics provides a direct link between the early state and
final observables. It has been extensively used in relativistic
heavy ion collisions. However, the usual hydrodynamic calculations
cannot explain the RHIC HBT data, $R_{\rm out} /R_{\rm side} \approx
1$, the so-called RHIC HBT puzzle \cite{hbtexp}.

In reference \cite{Zha04} a granular source model of QGP droplets
was put forth for the HBT puzzle. In references \cite{Zha06,Zha09}
the granular source model was improved to explain the RHIC HBT data
\cite{hbtexp}. In reference \cite{Won04}, the fluctuations of the
single-event two-pion correlation functions of granular sources were
discussed. In references \cite{Zha05,Ren08}, some observables for
granular sources were put forth and analyzed for the source
inhomogeneity in the heavy ion collisions at RHIC and Large Hadron
Collider (LHC) energies. The imaging characteristic quantities and
signals of the granular source of QGP droplets were presented in
references \cite{Zha09,Yan09}. This paper will give a review on the
important ingredients and recent progresses of the granular source
interferometry. \vspace*{-2mm}
\section{Granular source model of QGP droplets}
The granular structure of QGP droplets may occur in the first-order
phase transition from the QGP to hadronic gas, which was proposed by
E. Witten in 1984 \cite{Wit84}. In references \cite{Pra92,Zha95},
the two- and multi-pion HBT correlation functions of granular
sources were investigated.

In two-pion interferometry, there is the relation among the HBT
radii $R_{\rm out}$ and $R_{\rm side}$, the transverse velocity of
pion pair $v_T$, and source lifetime $\tau$, $R_{\rm out}^2 \approx
R_{\rm side}^2 + v_T^2 \tau^2$ \cite{Wie99}. In reference
\cite{Zha04}, it was noticed that the lifetime $\tau$ for a uniform
hydrodynamic evolving source scales with the source initial size.  A
smaller (or larger) initial source radius has a smaller (or larger)
$R_{\rm side}$, and smaller (or larger) source lifetime $\tau$. So,
there is always $R_{\rm out} > R_{\rm side}$ for the uniform
hydrodynamic sources. However, for a granular source with small QGP
droplets distributed in a large region, the source HBT radius
$R_{\rm out}$ is approximately equal to $R_{\rm side}$. It is
because that for the granular source the HBT radius $R_{\rm side}$
(proportional to the QGP droplets distribution) is much larger than
the source lifetime $\tau$ (proportional to the droplet radius).

In relativistic heavy ion collisions, the system initial transverse
energy density is highly fluctuating on event-by-event basis. The
large initial fluctuations together with the effects of violent
expansion and surface tension may lead to formation of granular
droplets \cite{Zha06,ZhaWon07}. On the other hand, the bulk
viscosity of the QGP may increase rapidly near the QCD transition,
which also leads to formation of the QGP droplets \cite{Tor08}.

It is assumed in the improved granular source model \cite{Zha09}
that the system produced in central relativistic heavy ion
collisions fragments and forms the QGP droplets at a time $t_0$
after $\tau_0$. The droplets distribute initially within a short
cylinder along the beam direction ($z$ direction) with the
probabilities
\begin{eqnarray}
\frac{dP_{\perp}}{2\pi\rho_0\,d\rho_0} \propto
\left[1-\exp\,(-\rho_0^2/\Delta{\cal
R}_{\perp}^2)\right]\theta({\cal R}_{\perp}-\rho_0)\,,
\end{eqnarray}
\vspace*{-5mm}
\begin{eqnarray}
\frac{dP_y}{dy_0}=\theta(y_m-|y_0|),~~~~z_0=t_0 \tanh y_0,
\end{eqnarray}
where $\rho_0$ and $z_0$ are the initial transverse and longitudinal
coordinates of the droplet center, $y_0$ is the initial rapidity of
the droplet, ${\cal R}_{\perp}$, $\Delta{\cal R}_{\perp}$, and $y_m$
are the radius of the cylinder, the shell parameter of the initial
distribution, and the central rapidity limitation. On the basis of
the Bjorken picture \cite{Bjo83}, the velocity of the droplet is
assumed as
\begin{equation}
\label{vdrop} v_{d \perp}=a_T \Big(\frac{\rho_0}{{\cal
R}_{\perp}}\Big)^{b_T}\sqrt{1-v_{dz}^2}\,,~~~~v_{dz}=z_0/t_0,
\end{equation}
where $a_T$ and $b_T$ are the magnitude and exponential power
parameters.

The evolution of the system after the fragmentation is the
superposition of all the evolutions of the individual droplets, each
of them is described by relativistic hydrodynamics with a cross-over
equation of state of the entropy density \cite{eos}. The initial
radius $r'_0$ of the droplets in local frame is assumed having a
Gaussian distribution with standard deviation $a$. To include the
pions emitted directly at hadronization and decayed from resonances
later, the pion freeze-out temperature is taken in a wide region
with the probability
\begin{eqnarray}
\label{Pt}
\frac{dP_f}{dT} \propto f_{\rm dir}
\exp[-(T-T_h)/\Delta T_{\rm dir}] + (1-f_{\rm dir}) \exp[-(T-T_h)/\Delta T_{\rm dec}]\,,
\end{eqnarray}
where $f_{\rm dir}$ is a fraction parameter for the direct emission,
$T_h$ is the temperature of complete hadronization, $\Delta T_{\rm
dir}$ and $\Delta T_{\rm dec}$ describe the widths of temperature
for the direct and decayed pion emissions.

The selections for the model parameter values can refer to
\cite{Zha09}.
\section{Granular source imaging}
The imaging technique introduced by Brown, Danielewicz, and Pratt
\cite{image} is a model-independent way to extract the two-pion
source function $S(r)$, the probability for emitting a pion pair
with spatial separation $r$ in the pair center-of-mass system
(PCMS), from the HBT correlation function.

\includegraphics{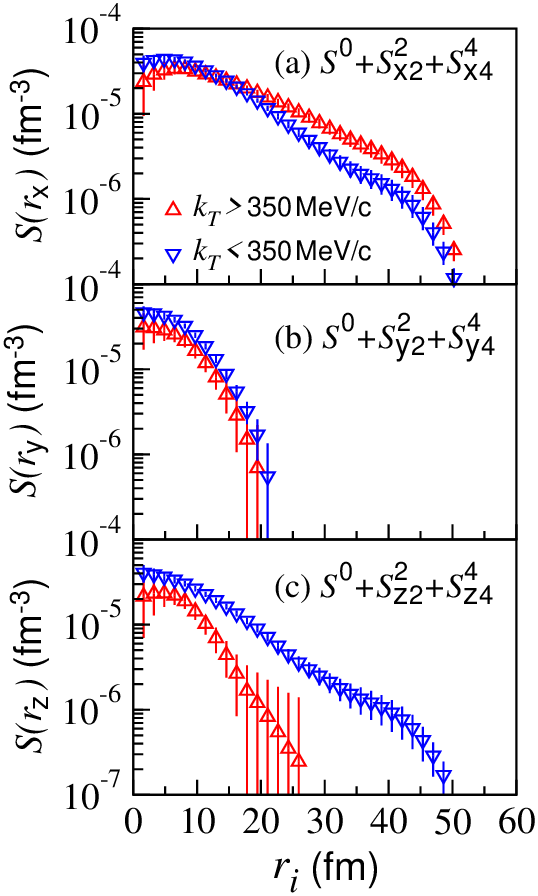} \vspace*{10.5cm} \vspace*{-3.3cm} \hangafter=0
\hangindent=-8.2cm \noindent Fig. 1. Three-dimension source
functions of the granular source.

\vspace*{-8.6cm} \includegraphics{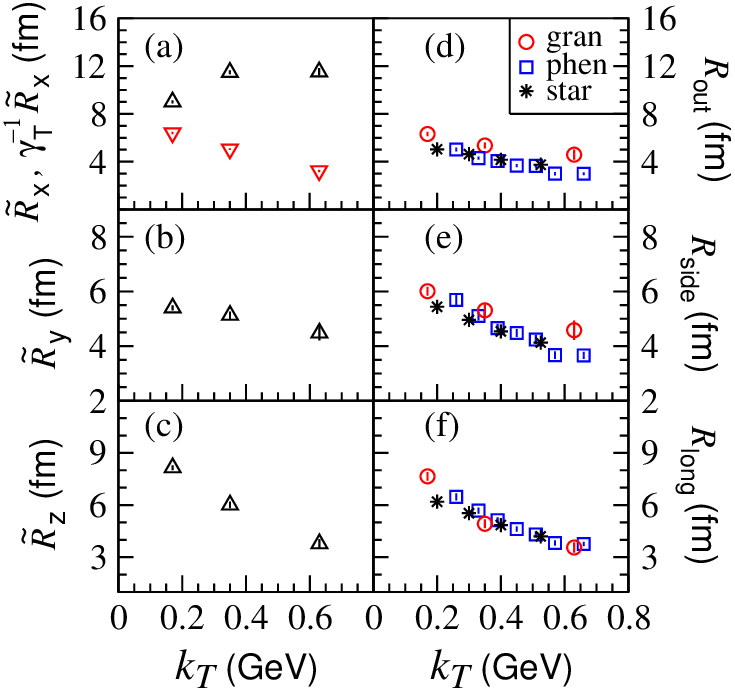} \vspace*{10.5cm} \vspace*{-3.4cm}
\hangafter=0 \hangindent=6.5cm \noindent Fig. 2. (a)--(d) Imaging
results of granular source. (e)--(f) HBT radii of granular source
and RHIC experiments \cite{hbtexp}.

\vspace*{3mm}Figure 1 (a), (b), and (c) show the source functions of
the granular source in $x$ (out), $y$ (side), and $z$ (long)
directions. Here $k_T$ is the transverse momentum of pion pair in
the longitudinally comoving system (LCMS). One can see that the
source functions in out and long directions have long tails. The
reason is the source expansion which boosts the pion pair in out and
long directions. The source function width in long direction is
smaller for larger $k_T$ because the average longitudinal momentum
of the pairs is smaller for larger $k_T$.

Once obtain the source functions one can introduce the quantities of
$r$ moments to describe the source geometry numerically. They are
model-independent quantities. Figure 2 (a)--(c) exhibit the
quantities, $\widetilde{R}_i\!=\!\!\sqrt{\pi}\langle r_i \rangle/2$,
$(i=x,y,z)$, of the first-order moment $\langle r_i \rangle$ for the
granular source. $\widetilde{R}_i$ describes the source size in $i$
direction and normalized to the Gaussian radius for one-dimension
Gaussian source \cite{Zha09}. For comparing, the usual Gaussian
fitted HBT radii of the granular source and RHIC experiments are
shown in Fig. 2 (d)--(f). In Fig. 2 (a), the symbols
{\raisebox{-1.2mm}{\textsuperscript{ $\nabla$}}} denote the results
of $\gamma^{-1}_T\widetilde{R}_x$, where $\gamma^{-1}_T$ is the
Lorentz contracted factor of LCMS to PCMS. One can see that the
imaging results of $\widetilde{R}_i$ are consistent with the HBT
radii after considering the Lorentz contraction.
\section{Granular source signals}
For granular sources, the single-event HBT correlation functions
exhibit large event-by-event fluctuations \cite{Won04,Zha05,Ren08}.
In order to observe the event-by-event fluctuations, we introduce
the quantity $f$ of the relative fluctuation of the single-event
correlation function to mixed-event correlation function, with its
error-inverses weight \cite{Zha05}

\includegraphics{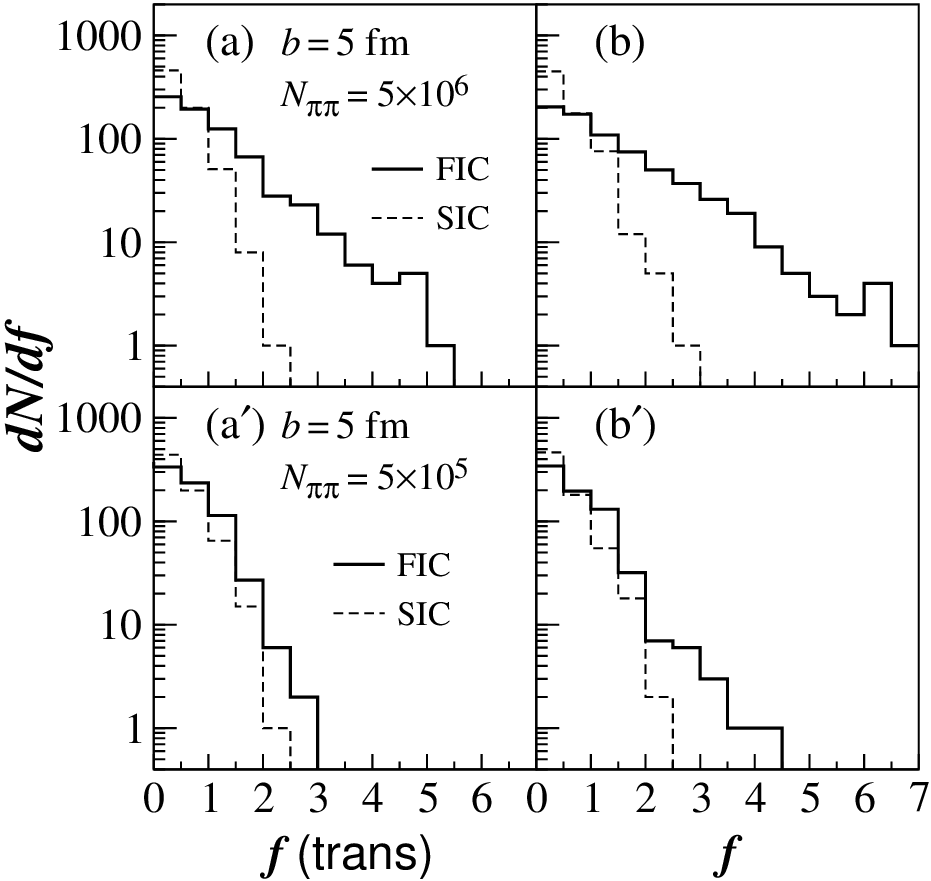} \vspace*{10.5cm} \vspace*{-4.0cm} \hangafter=0
\hangindent=-6.9cm \noindent Fig. 3. The distributions $dN/df$ for
40 events with FIC and SIC.

\vspace*{-8.6cm} \includegraphics{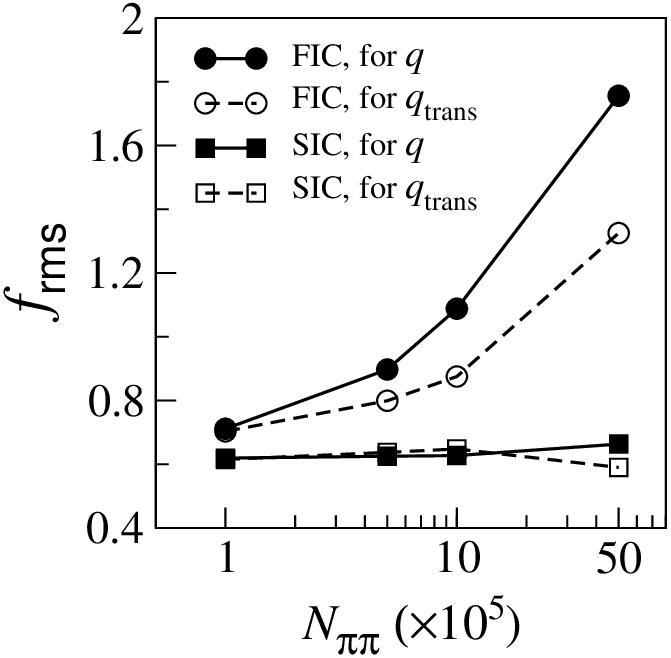} \vspace*{10.5cm} \vspace*{-2.9cm}
\hangafter=0 \hangindent=7.6cm \noindent Fig. 4. The
root-mean-square of $f$ as a function of $N_{\pi\pi}$.

\begin{equation}\label{RF}
f(q_i)=\frac{|C_s(q_i)-C_m(q_i)|}{\Delta |C_s(q_i)-C_m(q_i)|}.
\end{equation}
Figure 4 shows the distributions of the $f$ for the variables of
transverse relative momentum $q_{\rm trans}$ and relative momentum
$q$ of the pion pairs for the 40 events generated by the smoothed
particle hydrodynamics \cite{Ren08,Agu01}, with the impact parameter
$b=5$ fm. It can be seen that for the pion pair number
$N_{\pi\pi}=5\times10^6$, the distributions for the fluctuating
initial conditions (FIC) are much wider than those for the smoothed
initial conditions (SIC) both for $q_{\rm trans}$ and $q$. Even for
$N_{\pi\pi}=5\times10^5$, the widths for FIC are visibly larger than
those for SIC. Figure 5 shows the root-mean-square (RMS) of $f$, as
a function of $N_{\pi\pi}$ for the events. It can been seen that the
values of $f_{\rm rms}$ rapidly increase with $N_{\pi\pi}$ for FIC
because the errors in Eq. (\ref{RF}) decrease with $N_{\pi\pi}$. For
SIC the values of $f_{\rm rms}$ are almost independent from
$N_{\pi\pi}$, because both the differences and their errors in Eq.
(\ref{RF}) decrease with $N_{\pi\pi}$ in the case. At LHC energy the
event multiplicity of identical pions is about two thousands and the
order of $N_{\pi\pi}$ will be $10^6$. In this case, the
distributions of $f$ and its RMS will provide observable signals for
the source inhomogeneity.
\section{Discussion and conclusion}
In relativistic heavy ion collisions, the fluctuating initial
density distribution, violent expansion, surface tension, and
viscosity of the QGP may lead to formation of the granular sources.
For the granular sources, the single-event HBT correlation functions
have large event-by-event fluctuations. Because of data statistics
the experimental HBT analyses are performed with the mixed-event HBT
correlation functions, and the fluctuations are smoothed out after
the event average. However, the short lifetime of the granular
source may survive and lead to the HBT puzzle. The model-independent
quantities $\widetilde{R}_i$ obtained by imaging analysis for the
granular sources are consistent with the usual HBT radii of the
granular sources and RHIC experiments. At the coming LHC heavy ion
collisions, it is hopefully to observe the signals of the granular
sources.\\[2ex]
{\bf \large Acknowledgment}\\
This work was supported by the National Natural Science Foundation
of China under Grant No. 10775024.

\end{document}